\documentclass[aps,prl,twocolumn,superscriptaddress,amsmath,amssymb]{revtex4}
\usepackage{graphicx}

\begin{document}

\title{Stability Condition of a Strongly Interacting Boson-Fermion Mixture across an Inter-Species Feshbach Resonance}
\author{Zeng-Qiang Yu}
\affiliation{Institute for Advanced Study, Tsinghua University, Beijing, 100084, China}
\author{Shizhong Zhang}
\affiliation{Department of Physics, The Ohio-State University, Columbus, OH, 43210, USA}
\author{Hui Zhai}
\affiliation{Institute for Advanced Study, Tsinghua University, Beijing, 100084, China}
\date{\today}

\begin{abstract}
We study the properties of dilute bosons immersed in a single component Fermi sea across a broad boson-fermion Feshbach resonance. The stability of the mixture requires that the bare interaction between bosons exceeds a critical value, which is a universal function of the boson-fermion scattering length, and exhibits a maximum in the unitary region. We calculate the quantum depletion, momentum distribution and the boson contact parameter across the resonance. The transition from condensate to molecular Fermi gas is also discussed.

\end{abstract}
\maketitle

Feshbach resonances are an important tool for achieving strong interactions in degenerate quantum gases. The many-body system at resonance, if stable, is of great interest since the $s$-wave scattering length $a_{\text{s}}$ diverges and the system cannot be treated by conventional perturbative means.  On the other hand, universal properties arise at resonance and greatly simplify the problem. However, in order for a degenerate atomic gas to be stable, two conditions are required. First, the atom loss rate should be small to ensure long enough lifetime and second, the system should have a positive compressibility against mechanical collapse.

So far, two-component Fermi gases are the only systems whose stability at resonance has been firmly established and their properties have been studied extensively in the last  few years \cite{review}.
In contrast, bosons at resonance suffer from rapid atom loss \cite{boson-1} and collapse instability \cite{boson-2}, and are not stable.

For a boson-fermion mixture, the atom loss rate (dominated by three-body recombination) depends on the concentrations of the two species. In typical situations\cite{bf_exp}, the boson density $n_{\text{B}}$ is much larger than the fermion density $n_{\text{F}}$ and thus atom loss is also significant at resonance. However, in the limit of low boson concentration, most three-body processes involve two identical fermions and one boson, and the loss rate can be greatly suppressed. As for stability against collapse, the weak coupling mean-field theory for a uniform mixture gives
$k_{\text{F}}a_{\text{BB}}\geqslant (k_{\text{F}}a_{\text{BF}})^2(1+\gamma)^2/(2\pi\gamma)$  \cite{viverit},
where $k_{\text{F}}=(6\pi^2 n_{\text{F}})^{1/3}$, $\gamma=m_{\text{B}}/m_{\text{F}}$, $a_{\text{BB}}$ and $a_{\text{BF}}$ are the scattering lengths between bosons, and between bosons and fermions, respectively. In addition to the $^6$Li-$^{23}$Na and $^{40}$K-$^{87}$Rb mixtures studied previously \cite{bf_resonance}, new mixtures have been very recently been realized in the laboratory, including $^6$Li-$^{87/85}$Rb\cite{Li-Rb} and $^{40}$K-$^{41}$K \cite{Martin}.  It is therefore very important to generalize the stability condition of boson-fermion mixtures to the strongly interacting region where $a_{\text{BF}}\rightarrow \pm\infty$.

There have been a number of theoretical studies of boson-fermion mixtures.  Most address the region away from unitarity \cite{bf_theory_1,bf_theory_Giorgini} or deal with resonance physics using a mean-field treatment of a two-channel model \cite{bf_theory_2}. Some recent works studied boson-fermion pairing effects within a single-channel model for a broad resonance \cite{Fei}. However, the interaction between bosons, which is crucial for the stability of the mixture as we will show, is ignored in most of these analyses, and thus the question of stability is still an open one.

\begin{figure}
\includegraphics[width=8cm]{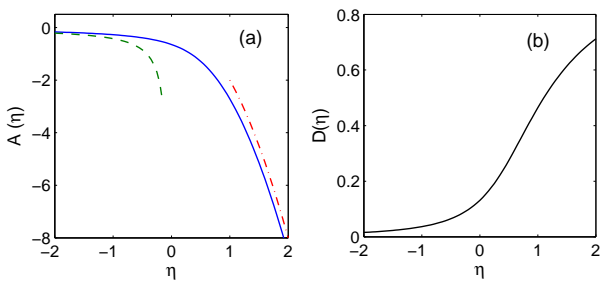}
\caption{(Color online) Solid line: Universal function $A(\eta)$ and $D(\eta)$ as a function of coupling constant $\eta=1/(k_\text{F} a_{\text{BF}})$. The dashed and dash-dotted lines in (a) are the mean-field value and molecular binding energy in the two limits, respectively.}\label{fig-energy}
\end{figure}

The Hamiltonian $\mathcal{H}$ for a boson-fermion mixture can be separated into $\mathcal{H}_0$ and $\mathcal{H}_1$ ($\hbar=1$),
\begin{align}
&\mathcal{H}_0=-\sum_{i=1}^{N_{\text{B}}}{\nabla_i^2\over 2m_{\text{B}}}-\sum_{j=1}^{N_{\text{F}}} {\nabla_j^2\over 2m_{\text{F}}}+\sum_{i=1}^{N_{\text{B}}}\sum_{j=1}^{N_{\text{F}}} U_{\text{BF}}({\bf r}_i^b-{\bf r}_j^f),\nonumber\\
&\mathcal{H}_1={1\over 2}\sum_{i,i^\prime=1}^{N_{\text{B}}} U_{\text{BB}}({\bf r}_i^b-{\bf r}_{i'}^{b}),
\end{align}
where $U_{\text{BF}}$ and $U_{\text{BB}}$ are the zero-range pseudo-potentials that produce the scattering lengths $a_{\text{BF}}$ and $a_{\text{BB}}$. In this work we consider the situation where $a_{\text{BF}}$ can be tuned by a broad Feshbach resonance and the boson-fermion interaction energy can be of the same order as the Fermi energy, and hence cannot be treated perturbatively. In this work we will use a Jastrow-Slater variational wave-function and the lowest order constrained variational (LOCV) approximation to study the ground state properties of $\mathcal{H}_0$.  On the other hand, we shall restrict ourselves in the regime $x\equiv n_{\text{B}}/n_{\text{F}}\ll 1$ so that $n^{1/3}_{\text{B}}a_{\text{BB}}$ is always small and positive. Therefore we can treat $\mathcal{H}_1$ as a perturbation and evaluate it at the mean-field level. Two main results of this Rapid Communication are summarized as follows:

(i) The energy density of the mixture is found to be given by
\begin{equation}
\frac{E}{V}=\frac{3}{5}\epsilon_\text{F}n_\text{F}+\frac{1+\gamma}{2\gamma}A(\eta)n_\text{B}\epsilon_\text{F}+\frac{1}{2}g_\text{BB}n^2_\text{B}\big[1+D(\eta)\big],\label{Energy}
\end{equation}
where $\epsilon_{\text{F}}= k^2_{\text{F}}/(2m_{\rm F})$, $g_{\rm BB}=4\pi a_{\rm BB}/m_{\rm B}$ and $\eta=1/(k_\text{F}a_\text{BF})$. $A(\eta)$ and $D(\eta)$ are two universal functions of $\eta$ and are shown in Fig. \ref{fig-energy}. These two functions have simple physical interpretations. In the boson-fermion mixture, minority bosons are dressed by a polarized cloud of fermions and form bosonic polarons; $A(\eta)\epsilon_{\rm F}$ is the polaron binding energy for $\gamma=1$. $D(\eta)$ is the fraction of non-condensed bosons (quantum depletion) arising from boson-fermion interactions.

\begin{figure}[tbp]
\includegraphics[width=8cm]{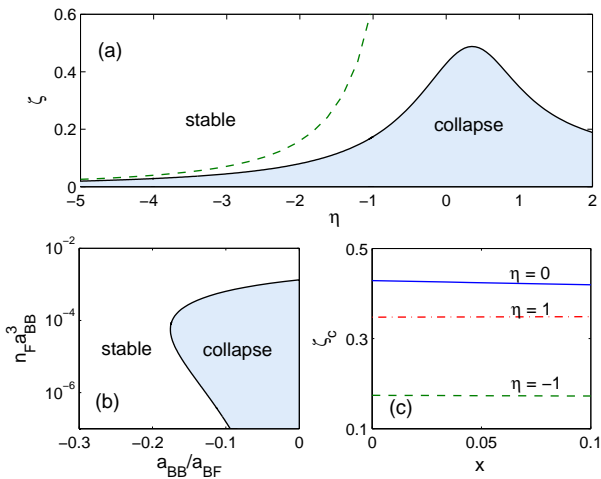}
\caption{(Color online) Phase diagram of dilute bosons immersed in a single component Fermi gas: (a) in terms of $\eta=1/(k_{\text{F}}a_{\text{BF}})$ and $\zeta=k_{\text{F}}a_{\text{BB}}$; (b) in terms of $n_{\text{F}}a^3_{\text{BB}}$ and $a_{\text{BB}}/a_{\text{BF}}$. The dashed line in (a) is the mean-field result in the weak coupling limit. (c) Critical value $\zeta_\text{c}$ as a function of boson concentration $x=n_\text{B}/n_\text{F}$ for different $\eta$. For all cases we set $m_{\rm B}=m_{\rm F}$. }\label{fig-zeta_c}
\end{figure}

(ii) With the ground state energy in Eq. (\ref{Energy}) we determine the mechanical stability condition,
\begin{equation}
k_{\text{F}}a_{\text{BB}}\geqslant \zeta_\text{c}(\eta, \gamma,x),\label{stability}
\end{equation}
across a Feshbach resonance. In the dilute limit $x\ll 1$, to zeroth order in $x$,
\begin{equation}
\zeta_\text{c}=\frac{\pi(1+\gamma)^2}{32\gamma\big(1+D(\eta)\big)}\left(2A(\eta)-\eta\frac{\partial A(\eta)}{\partial \eta}\right)^2.\label{zeta_c}
\end{equation}
Here $\zeta_\text{c}$ is a universal function of $\eta$ for given $\gamma$ as shown in Fig. \ref{fig-zeta_c}(a).  It attains a value of the order unity at resonance and is maximum in the unitarity region.

{\it Jastrow-Slater Calculation of Energy Density.} We use a variational wave-function
\begin{align}
 |\Psi_\text{JS}\rangle =\prod_{i,j} f({\bf r}_i^b-{\bf r}_j^f) \left(\frac{1}{\sqrt{V}}\right)^{N_\text{B}}|\Phi_{\rm FS}\rangle, \label{jastrow}
\end{align}
where $|\Phi_{\text{FS}}\rangle$ represents a Fermi sea with $N_\text{F}$ fermions and $f({\bf r})$ is the Jastrow function describing the two-body correlations between bosons and fermions. $f({\bf r})$ deviates from unity only within the so-called healing length $d$, i.e., $f(r>d)=1$. To estimate the energy $E_0=\langle \Psi_\text{JS}|\mathcal{H}_0|\Psi_\text{JS}\rangle/ \langle \Psi_\text{JS}|\Psi_\text{JS}\rangle$, we use the so-called LOCV  approximation \cite{LOCV}, which basically treats correlations to lowest order (${\cal{O}}(f^2-1)$) in a linked cluster expansion. LOCV was first used to study $^4$He \cite{LOCV}, and has also recently been applied to study bosons with large scattering lengths \cite{pethick} and Fermi gases at resonance \cite{thesis}. The calculation is both simpler and more transparent than Monte-Carlo simulations while providing a fair approximation to Monte-Carlo results.

Within LOCV, we find $E_0/V=\mathcal{E}_{\text{F}}^0+\mathcal{E}_{\text{BF}}$, where $\mathcal{E}_\text{F}^0\equiv 3 n_{\text{F}}\epsilon_{\text{F}}/5$ is the energy of the free fermions and $\mathcal{E}_{\text{BF}}$ is the interaction energy
\begin{align}
\mathcal{E}_{{\rm BF}}=n_\text{B}n_\text{F}\int {\rm d}^3{\bf r}\, f({\bf r})\left[-{\nabla_{\bf r}^2\over 2m_\text{r}} + U_{\text{BF}}({\bf r})\right]f({\bf r}) \label{EpLOCV}
\end{align}
between bosons and fermions with $m_\text{r}=m_\text{B}m_\text{F}/(m_\text{B}+m_\text{F})$. Eq. (\ref{EpLOCV}) can be rewritten as $\mathcal{E}_\text{BF}= \lambda n_\text{B}=A(\eta,\gamma)n_\text{B}\epsilon_\text{F}$, where $\lambda$ is found by solving $\big[-{1\over2m_\text{r}}{{\rm d}^2\over {\rm d}r^2}+U_{\text{BF}}(r)\big]rf(r)=\lambda rf(r)$, subject to the constraint $4\pi n_\text{F} \int_0^d {\rm d}r\, r^2|f(r)|^2=1$ and the boundary conditions $\left.{(rf)'\over rf}\right|_{r=0}=-{1\over a_{\text{BF}}}$, $f(d)=1$ and $ \left.f(r)'\right|_{r=d}=0$. It follows that the universal function $A$ is given by
$A(\eta,\gamma)=\lambda(\eta,\gamma)/\epsilon_F=(1+\gamma)A(\eta)/(2\gamma)$.
For a given value of $\eta$, there is only a change of prefactor for different mass ratios $\gamma$.

The numerical solution of $A(\eta)$ across the resonance is shown in Fig. \ref{fig-energy}(a). In the limit $\eta\rightarrow -\infty$, we find $A(\eta) =4/(3\pi\eta)$, so the interaction energy density reduces to the mean-field result $2\pi a_{\text{BF}}n_\text{F}n_{\text{B}}/m_\text{r}$. In the opposite limit $\eta\rightarrow +\infty$, $A(\eta) =-2\eta^2$, and $\lambda$ becomes the binding energy of the molecule $-1/(2m_\text{r}a_{\text{BF}}^2)$.

At resonance ($\eta=0$), we obtain $A=-0.64$ for equal masses. It is interesting to compare this result with related studies of highly polarized fermion-fermion mixtures, where minority fermions form polarons and the system behaves as a polaron Fermi liquid. Various theoretical approaches have been used to estimate the binding energy of a single impurity \cite{theory,chevy,QMC}. A variational approach gives $A=-0.61$ \cite{chevy} and diagrammatic Monte Carlo gives $A=-0.62$ \cite{QMC}. Experiments yield $A=-0.64(7)$ \cite{exp1}.

\begin{figure}[tbp]
\includegraphics[width=8.1 cm]{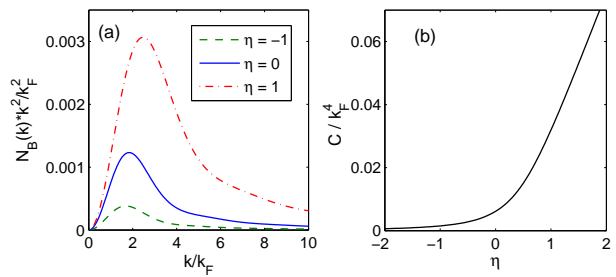}
\caption{(Color online) (a) Momentum distribution of non-condensed bosons $N_{\rm B}({\bf k})k^2/k^2_\text{F}$ for different $\eta=1/(k_\text{F}a_\text{BF})$; (b) Contact $\mathcal{C}$ (in unit of $k^{-4}_\text{F}$) as a function of $\eta$. For both cases we set $x=n_{\rm B}/n_{\rm F}=0.1$.} \label{fig-momentum}
\end{figure}

LOCV can also be used to determine the quantum depletion of bosons $n_\text{dep}$ as well as their momentum distribution $N_{\rm B}({\bf k})$:
\begin{align}
&D(\eta)={n_\text{dep}\over n_\text{B}}
 =n_\text{F} \int{\rm d}^3{\bf r}\, [f({\bf r})-1]^2,\\
&N_{\rm B}({\bf k}\neq 0)=n_\text{B} n_\text{F}\left|\int d^3{\bf r} \left(f({\bf r})-1\right)e^{-i{\bf k}\cdot{\bf r}}\right|^2
\end{align}
The depletion fraction $D(\eta)$ across the resonance is plotted in Fig. \ref{fig-energy}(b).  It is a monotonically increasing function of $\eta$. At resonance, we find $D=0.13$. Similar calculations have been carried out for a resonant Bose gas by Cowell {\it et al.} \cite{pethick}. The momentum distribution of depleted bosons are plotted in Fig. \ref{fig-momentum}(a) for different $\eta$, and the contact $\mathcal{C}=\lim_{k\rightarrow +\infty}k^4 N_{\rm B}({\bf k})$ \cite{Contact} as a function of $\eta$ is plotted in Fig. \ref{fig-momentum}(b). We note that at low momentum, $N_{\rm B}({\bf k})$ will be modified by boson-boson interactions. However, this will not affect our analysis of the large-${\bf k}$ limit of $N_{\rm B}({\bf k})$ or the stability analysis below.

With the solution for $|\Psi_{\rm JS}\rangle$, the energy correction due to $\mathcal{H}_1$ is obtained to lowest order:
\begin{align}
{E_1\over V}={1\over V}{\langle\Psi_{\text{JS}}|\mathcal{H}_1|\Psi_\text{JS}\rangle\over \langle\Psi_{\text{JS}}|\Psi_\text{JS}\rangle}=\frac{1}{2}g_\text{BB}n^2_\text{B}\big[1+4D(\eta)\big]. \label{E1}
\end{align}
Physically, the first term in Eq. (\ref{E1}) is the Hartree energy between the interacting bosons, and the second term can be interpreted as the Fock energy and pair hopping energy arising from interactions between condensed and non-condensed bosons.

{\it Stability Condition.} Using Eqs. (\ref{EpLOCV}) and (\ref{E1}),  Eq. (\ref{Energy}) gives the total energy of the system. In the weak coupling limit our theory agrees quantitatively with the perturbation results of Ref. \cite{bf_theory_Giorgini}.  From Eq. (\ref{Energy}) we obtain
\begin{align}
&\mu_\text{B}=\epsilon_\text{F}\left[\frac{1+\gamma}{2\gamma}A(\eta)+\frac{4\zeta}{3\pi\gamma}\big(1+D(\eta)\big)x\right]\nonumber\\
&\mu_\text{F}=\epsilon_\text{F}\left[1+\frac{1+\gamma}{6\gamma}\Big(2A(\eta)-\eta\frac{\partial A(\eta)}{\partial \eta}\Big)x-\frac{2\zeta\eta}{9\pi\gamma}\frac{\partial D(\eta)}
{\partial \eta}x^2\right]\nonumber
\end{align}
where $\zeta\equiv k_\text{F}a_\text{BB}$. The stability of the system requires (i) $\partial \mu_\text{F}/\partial n_\text{F}>0$; (ii) $\partial \mu_\text{B}/\partial n_\text{B}>0$ and (iii) the stability against small density fluctuation requires ${\rm Det} (M)>0$,
\begin{equation}
M=\left(\begin{array}{cc}\partial \mu_\text{F}/\partial n_\text{F} & \partial \mu_\text{F}/\partial n_\text{B} \\ \partial \mu_\text{B}/\partial n_\text{F} & \partial \mu_\text{B}/\partial n_\text{B}\end{array}\right)
\end{equation}
Here (i) and (ii) are easy to satisfy, while (iii) gives $\zeta>\zeta_\text{c}(\eta,\gamma,x)$, and $\zeta_{\rm c}$ is given by Eq. (\ref{zeta_c}) in the dilute limit $x\ll 1$.  For equal masses, the phase diagram in the dilute limit is shown in Fig. \ref{fig-zeta_c}. Physically, collapse is a global instability due to negative boson-fermion mean-field energy, and only sufficiently strong boson-boson repulsions prevent the density from increasing without bound.  In the $\eta-\zeta$ plane (Fig. \ref{fig-zeta_c}(a)), the phase boundary is given by the universal function $\zeta_\text{c}(\eta)$.
At resonance, we find $\zeta_\text{c}=0.43$. In the limit $\eta\rightarrow -\infty$, one finds $\partial A/\partial\eta= -A/\eta$ and $D(\eta)\rightarrow 0$.  Thus, $\zeta_\text{c}=(1+\gamma)^2/(2\pi\gamma \eta^2)$,
which agrees with the mean-field result $\zeta_\text{c}\rightarrow 0$ \cite{viverit}.
In the opposite limit, $\eta\rightarrow +\infty$,  one also finds $\zeta_c\rightarrow 0$ because $2A(\eta)-\eta \partial A(\eta)/\partial \eta
\rightarrow 0$. Therefore, a maximum of $\zeta_{\rm c}$ is expected in the unitary region.  For equal masses, the maximum of $\zeta_\text{c}$ is $0.49$,
above which the system is stable for all values of $\eta$. Physically, the non-monotonic behavior of $\zeta_\text{c}$ arises from the off-diagonal term $\partial \mu_\text{B}/\partial n_\text{F}$ in $M$. For $\eta\rightarrow -\infty$, $|\partial\mu_\text{B}/\partial n_\text{F}|$ vanishes linearly with $|a_\text{BF}|\rightarrow 0$;
while for $\eta\rightarrow +\infty$, the attraction becomes so strong that only short-range physics matters,
$\mu_\text{B}$ eventually approaches the molecular binding energy $-1/(2m_\text{r}a^2_\text{BF})$ independent of $n_\text{F}$,
and $|\partial \mu_\text{B}/\partial n_\text{F}|$ vanishes again. Hence $|\partial\mu_\text{B}/\partial n_\text{F}|$ must exhibit a maximum in between which gives rise to the maximum of $\zeta_\text{c}$.

Another notable feature of the phase diagram is that, for a given scattering length $a_{\rm BB}$, the stability requirement for $n_\text{F}$ is opposite in the weakly interacting and resonance regions.   For weak interactions, stability requires $n_{\text{F}}\leqslant \big[2\pi a_{\text{BB}}\gamma/(a^2_{\text{BF}}(1+\gamma)^2)\big]^3/(6\pi^2)$; while at resonance it requires $n_{\text{F}}\geqslant(\zeta_\text{c}/a_{\text{BB}})^3/(6\pi^2)$. In Fig. \ref{fig-zeta_c} (b), the phase diagram is plotted in terms of $n_\text{F}a^3_\text{BB}$ and $a_\text{BB}/a_\text{BF}$. For $|a_\text{BB}/a_\text{BF}|>0.18$, the system is stable for all values of $n_\text{F}$ and for $|a_\text{BB}/a_\text{BF}|<0.18$, as $n_\text{F}$ increases, the system first becomes unstable and then becomes stable again. In Fig.~\ref{fig-zeta_c}(c), we see that $\zeta_\text{c}$ depends weakly on the boson concentration $x$ for small $x$.

\begin{figure}[tbp]
\includegraphics[width=8.0cm]{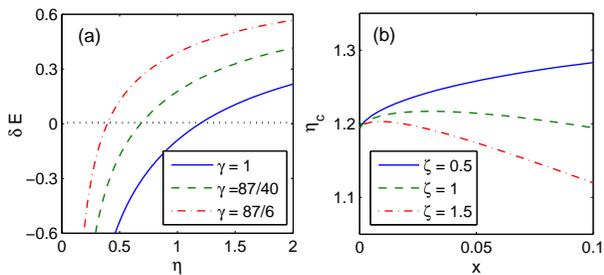}
\caption{(Color online) (a) $\delta E$, defined as the energy difference between the condensate [Eq. (\ref{Energy})] and atom-molecule mixture [Eq. (\ref{molecule-energy})] energies as a function of $\eta=1/(k_\text{F}a_\text{BF})$. We consider the case of equal masses as well as the mixtures of $\rm ^{87}$Rb-$\rm^{40}$K and $\rm ^{87}$Rb-$\rm^6$Li. (b) The estimated transition point $\eta_\text{c}$ as a function of $x=n_\text{B}/n_\text{F}$ for $\gamma=1$ and different $\zeta$.} \label{fig-molecule}
\end{figure}

{\it Applicable Region of Our Theory.} For $\eta\gg 1$, strong pairing fluctuations lead to the formation of fermionic molecules, and a Jastrow-Slater type wave-function is no longer applicable. The mean-field energy of a uniform atom-molecule mixture is given by \cite{seperation}
\begin{align}
  \frac{E_\text{M}}{V}
  = \mathcal{E}_{\rm F}^0\left[(1-x)^{5/3}+{1\over 1+\gamma}x^{5/3}-{5(1+\gamma)\eta^2\over 3\gamma}x \right. \nonumber\\
  \left. + {10(2+\gamma)\over 9\pi(1+\gamma)\eta}{a_{\text{MF}}\over a_{\text{BF}}}x(1-x)\right],
  \label{molecule-energy}
\end{align}
where $a_{\text{MF}}$ is the molecule-fermion scattering length \cite{Petrov2}. For imbalanced fermion-fermion mixtures, a similar mean-field energy has been found to be quite accurate compared with Monte Carlo results when $\eta>1$ \cite{zwerger}. Here we estimate the critical value $\eta_\text{c}$ of the phase boundary by comparing the energy in Eq. (\ref{molecule-energy}) with the energy Eq. (\ref{Energy}) discussed above.
To zeroth order in $x$, $\eta_\text{c}$ is determined by the equation
\begin{equation}
 A(\eta_\text{c},\gamma)=-1-{1+\gamma\over \gamma}\eta_\text{c}^2+{2(2+\gamma)\over 3\pi(1+\gamma)\eta_\text{c}}{a_{\text{MF}}\over a_{\text{BF}}}.
\end{equation}
The energy comparison is shown in Fig. \ref{fig-molecule}(a). For $\gamma=1$ we obtain $\eta_\text{c}=1.19$, and $\eta_\text{c}$ moves toward resonance as $\gamma$ increases.  Fig. \ref{fig-molecule}(b) shows that $\eta_{\rm c}$ depends weakly on $x$. We note that the energy comparison only gives a rough estimate of the phase boundary. The nature of the transition, i.e.  whether the loss of condensation and the appearance of a molecular Fermi surface occur at one first-order critical point or two separated second-order critical points, is left for future work.

{\it Experimental Realization.} In this work, we analyze the stability condition and determine the phase diagram of a boson-fermion mixture across a Feshbach resonance. Experimentally, at a given magnetic field near the boson-fermion resonance, $a_{\text{BB}}$ is fixed. However, if one combines other control techniques, for instance, using optical Feshbach or microwave-induced resonances, one can also tune $a_{\text{BB}}$  independently as proposed recently in Refs. \cite{peng}. The mechanical stability condition Eq. (\ref{stability}) predicted here will provide a useful guide for experiments searching for stable, strongly-interacting boson-fermion mixtures.  The depletion fraction, momentum distribution and contact calculated here can be measured by various techniques such as Bragg spectroscopy \cite{Bragg} and radio-frequency spectroscopy \cite{ARPES}.

{\it Acknowledgements.} We thank Tin-Lun Ho, Ran Qi, Zhenhua Yu, Xiaoling Cui, Peng Zhang and Edward Taylor for helpful discussions and reading the manuscript. This work is supported by Tsinghua University Initiative Scientific Research Program, NSFC Grant No. 11004118, and NKBRSFC under Grants No. 2011CB921500.


\begin{thebibliography}{99}




\bibitem{review}
S. Giorgini, L. P. Pitaevskii, and S. Stringari, Rev. Mod. Phys. {\bf 80}, 1215 (2008).


\bibitem{boson-1}
J. Stenger, {\it et al}., Phys. Rev. Lett. {\bf 82}, 2422 (1999).

\bibitem{boson-2}
S. Basu and E. J. Mueller, Phys. Rev. A {\bf 78}, 053603 (2008)

\bibitem{bf_exp}
G. Modugno, {\it et al}., Science {\bf 297}, 2240 (2002);
C. Ospelkaus, S. Ospelkaus, K. Sengstock, and K. Bongs, Phys. Rev. Lett. {\bf 96}, 020401 (2006).


\bibitem{viverit}
L. Viverit, C. J. Pethick, and H. Smith, Phys. Rev. A {\bf 61}, 053605 (2000).

\bibitem{bf_resonance}
C. A. Stan, M. W. Zwierlein, C. H. Schunck, S. M. F. Raupach, and W. Ketterle, Phys. Rev. Lett. {\bf 93}, 143001 (2004);
S. Inouye, {\it et al}., {\it ibid}. {\bf 93}, 183201 (2004);
S. Ospelkaus, C. Ospelkaus, L. Humbert, K. Sengstock, and K. Bongs,  {\it ibid}. {\bf 97}, 120403 (2006);
F. Ferlaino, {\it et al}., Phys. Rev. A {\bf 73}, 040702(R) (2006);
M. Zaccanti, C. D¡¯Errico, F. Ferlaino, G. Roati, M. Inguscio, and G. Modugno, {\it ibid}. {\bf 74}, 041605(R) (2006).

\bibitem{Li-Rb}
B. Deh, C. Marzok, C. Zimmermann, and P. W. Courteille, Phys. Rev. A {\bf 77}, 010701(R) (2008);
B. Deh, {\it et al}., {\it ibid}. {\bf 82}, 020701(R) (2010).

\bibitem{Martin}
C.-H. Wu, {\it et al}., arXiv:1103.4630.

\bibitem{bf_theory_1}
A. P. Albus, S. A. Gardiner, F. Illuminati, and M. Wilkens, Phys. Rev. A {\bf 65}, 053607 (2002);
D. H. Santamore, S. Gaudio, and E. Timmermans, Phys. Rev. Lett. {\bf 93}, 250402 (2004);

\bibitem{bf_theory_Giorgini}
L. Viverit and S. Giorgini, Phys. Rev. A {\bf 66}, 063604 (2002);

\bibitem{bf_theory_2}
S. Powell, S. Sachdev, and H. P. B\"{u}chler, Phys. Rev. B {\bf 72}, 024534 (2005);
F. M. Marchetti, C. J. M. Mathy, D. A. Huse, and M. M. Parish, {\it ibid}. {\bf 78}, 134517 (2008).

\bibitem{Fei}
J. L. Song, M. S. Mashayekhi, and F. Zhou, Phys. Rev. Lett. {\bf 105}, 195301 (2010); E. Fratini and P. Pieri, Phys. Rev. A {\bf 81}, 051605(R) (2010); and T. Watanabe, T. Suzuki, and P. Schuck, Phys. Rev. A {\bf 78}, 033601 (2008);


\bibitem{LOCV}
V. R. Pandharipande, Nucl. Phys. A {\bf 174}, 641 (1971); {\it ibid}. {\bf 178}, 123 (1971); V. R. Pandharipande and H. A. Bethe, Phys. Rev. C {\bf 7}, 1312 (1973).

\bibitem{pethick}
S. Cowell, {\it et al}., 
Phys. Rev. Lett. {\bf 88}, 210403 (2002).

\bibitem{thesis}
S. Y. Chang, PhD thesis, 2006, University of Illinois at Urbana-Champaign.

\bibitem{theory}
C. Lobo, A. Recati, S. Giorgini, and S. Stringari, Phys. Rev. Lett. {\bf 97}, 200403 (2006);
R. Combescot, A. Recati, C. Lobo, F. Chevy, {\it ibid}. {\bf 98}, 180402 (2007);
R. Combescot, and S. Giraud, {\it ibid}. {\bf 101}, 050404 (2008).

\bibitem{chevy}
F. Chevy, Phys. Rev. A {\bf 74}, 063628 (2006).

\bibitem{QMC}
N. Prokof'ev and B. Svistunov, Phys. Rev. B {\bf 77}, 020408(R) (2008); {\it ibid}. {\bf 77}, 125101 (2008).

\bibitem{exp1}
A. Schirotzek, {\it et al.},
Phys. Rev. Lett. {\bf 102}, 230402 (2009). S. Nascimb\`ene, {\it et al.}, {\it ibid}. {\bf 103}, 170402 (2009).



\bibitem{Contact}
S. Tan, Ann. Phys., {\bf 323}, 2971 (2008); E. Braaten and L. Platter, Phys. Rev. Lett. {\bf 100}, 205301 (2008); S. Zhang and A. J. Leggett, Phys. Rev. A {\bf 79}, 023601 (2009).

\bibitem{seperation}
We have verified that for dilute molecules, the repulsive interaction between molecules and fermionic atoms will not lead to phase separation within mean-field theory.
\bibitem{Petrov2}
D. S. Petrov, Phys. Rev. A {\bf 67}, 010703(R) (2003).

\bibitem{zwerger}
M. Punk, P. T. Dumitrescu, and W. Zwerger, Phys. Rev. A {\bf 80}, 053605 (2009).

\bibitem{peng}
P. Zhang, P. Naidon, and M. Ueda, Phys. Rev. Lett. {\bf 103}, 133202 (2009); T. M. Hanna, E. Tiesinga, P. S. Julienne, New J. Phys. {\bf 12}, 083031 (2010); D. J. Papoular, G. V. Shlyapnikov, and J. Dalibard, Phys. Rev. A {\bf 81}, 041603 (2010); T. V. Tscherbul, {\it et al.},
{\it ibid}. {\bf 81}, 050701 (2010).

\bibitem{Bragg}
E. D. Kuhnle, {\it et al}., Phys. Rev. Lett. {\bf 105}, 070402 (2010).

\bibitem{ARPES}
J. T. Stewart, {\it et al}., 
Phys. Rev. Lett. {\bf 104}, 235301 (2010).




\end{thebibliography}
\end{document}